\documentclass[3p,twocolumn,authoryear]{elsarticle}

 \usepackage{amsmath} 
\usepackage{amssymb} 
\usepackage{mathrsfs} 
\usepackage{psfrag}
\usepackage{amsfonts}
\usepackage{booktabs}

\usepackage{doi}
\usepackage[usenames,dvipsnames]{xcolor}
\usepackage{hyperref}
\hypersetup{
        breaklinks,
        colorlinks=true,
        linkcolor=blue,
        citecolor=green,
        pdfstartview={FitH},
        pdfview={FitH 0},
        pdfauthor={T J W Wagner},
        pdftitle={On the representation of capsizing in iceberg models},
 }

\newcommand{\be}[0]{\begin{equation}}
\newcommand{\ee}[0]{\end{equation}}

\newcommand{\dx}{\Delta x}
\newcommand{\ri}{\rho_i}
\newcommand{\rw}{\rho_w}

\newcommand{\vi}{\vec{v}_i}
\newcommand{\vw}{\vec{v}_w}
\newcommand{\va}{\vec{v}_a}

\journal{arXiv}

\begin{document}

\begin{frontmatter}



\title{On the representation of capsizing in iceberg models}


\author[SIO]{Till J.W. Wagner\corref{corr1}}
\cortext[corr1]{email: tjwagner@ucsd.edu}
\address[SIO]{Scripps Institution of Oceanography, University of California San Diego}

\author[PRI]{Alon A. Stern}
\address[PRI]{Geophysical Fluid Dynamics Laboratory, Princeton University}

\author[SIO]{Rebecca W. Dell}

\author[SIO]{Ian Eisenman}

\begin{abstract}
Although iceberg models have been used for decades, they have received far more widespread attention in recent years, in part due to efforts to explicitly represent icebergs in climate models. This calls for increased scrutiny of all aspects of typical iceberg models. An important component of iceberg models is the representation of iceberg capsizing, or rolling. Rolling occurs spontaneously when the ratio of iceberg width to height falls below a critical threshold. Here we examine previously proposed representations of this threshold, and we find that there have been crucial errors in the representation of rolling in many modeling studies to date. We correct these errors and identify an accurate model representation of iceberg rolling. Next, we assess how iceberg rolling influences simulation results in a hierarchy of models. Rolling is found to substantially prolong the lifespan of individual icebergs and allow them to drift farther offshore, although it is found to have relatively small impacts on the large-scale freshwater distribution in comprehensive model simulations. The results suggest that accurate representations of iceberg rolling may be of particular importance for operational forecast models of iceberg drift, as well as for regional changes in high-resolution climate model simulations.
\end{abstract}


\end{frontmatter}


\section{Introduction} \label{sec:intro}
 
The drift and decay of icebergs has received increasing interest in recent years associated with several factors. \emph{(i)}  Icebergs pose a threat to high-latitude shipping, as well as to offshore hydrocarbon exploration efforts. The rapid retreat of Arctic sea ice and concurrent increases in oil and gas demands have increased shipping through the Arctic \citep{Pizzolato:2014cy} and discussions of drilling operations in the Arctic Ocean \citep{Unger:2014uv,Henderson:2016wa}. \emph{(ii)}  Global warming, and particularly high temperatures observed around Greenland and the Antarctic Peninsula, are being linked to increases in the flux of icebergs calving from glaciers and ice shelves. Calving rates are thus projected to accelerate during the coming decades \citep[e.g.,][]{Rignot:2006fm, Copland:2007du, Rignot:2011hi,Joughin:2014ew}. \emph{(iii)} An increased incidence of icebergs is projected to impact regional ecosystems and oceanographic conditions \citep[e.g.,][]{Vernet:2012th,Smith:2013cu,Stern:2015bo,Duprat:2016hw}. 
\emph{(iv)} Icebergs carry and release freshwater far from the calving source, and they can affect the large-scale ocean circulation \citep[e.g.,][]{Martin:2010kb, Stern:2016kh};  
\emph{(v)} As an extreme example of this, the release of massive armadas of icebergs from the Laurentide Ice Sheets during the Heinrich Events of the last glacial period are believed to have affected Earth's climate globally \citep[see e.g.,][]{Broecker:1994em, Stokes:2015dt}.
In light of these factors, icebergs have recently begun to be explicitly represented in state-of-the-art comprehensive global climate models (GCMs) \citep[e.g.,][]{Martin:2010kb, Hunke:2011fx, Stern:2016kh}. 

 \begin{figure*}[ht!]
 \begin{center}
 \hspace{-.5 cm} \includegraphics[width=\linewidth]{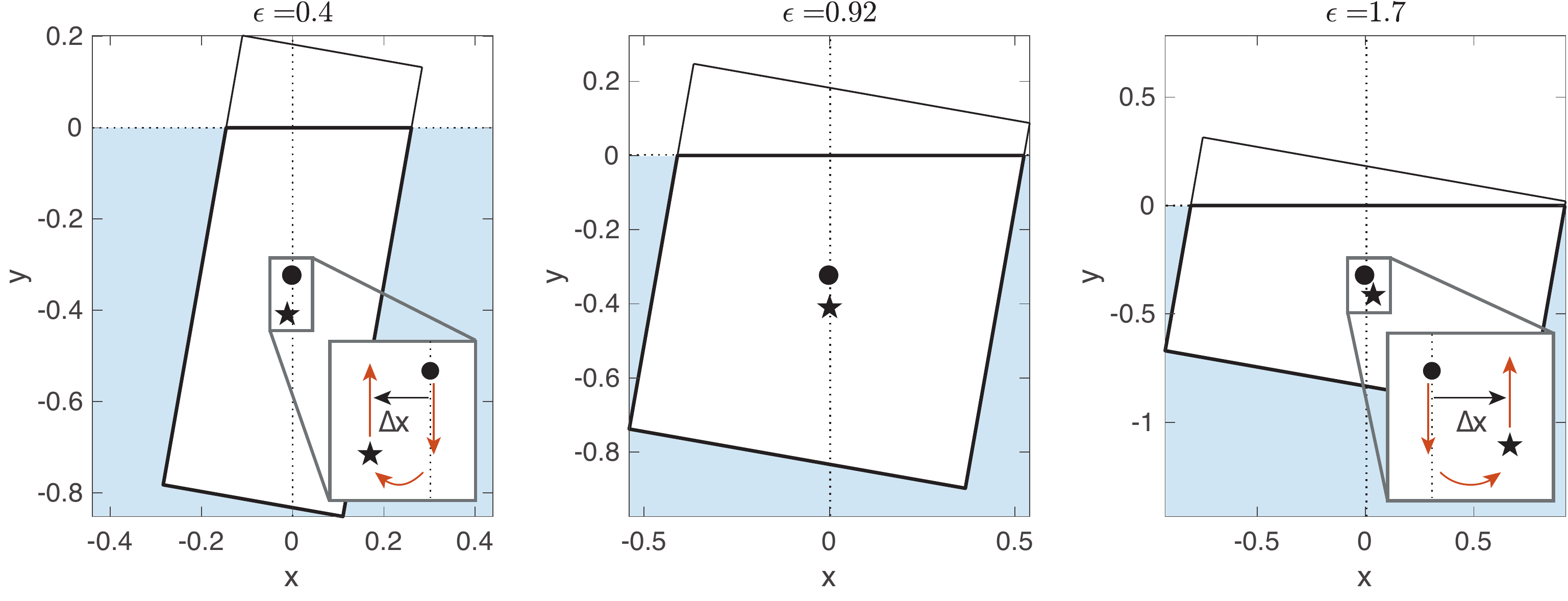}
 \caption{Schematics of free-floating icebergs with various aspect ratios, $\epsilon \equiv W/H$. Here, $\alpha = 0.83$ and $\theta = 10^\circ$. Shown are the center of gravity ({\normalsize$\bullet$}) and center of buoyancy ({\scriptsize$\bigstar$}).  (a) $\epsilon = 0.4$, with $\dx<0$, giving an unstable iceberg that will roll over. (b) Marginally unstable iceberg with $\epsilon = \epsilon_c = 0.92$. In the present case, with $\theta = 10^\circ$, $\dx = -0.01$. In the limit $\theta \rightarrow 0$, $\dx(\epsilon_c) = 0$. (c) Stable, self-righting iceberg, with $\epsilon = 1.7$ and $\dx>0$. Insets indicate the horizontal shift of the center of buoyancy, $\dx$. The red arrows illustrate gravity, buoyancy and resulting torque (insets).}
 \label{fig:bergschem}
 \end{center}
\end{figure*}

The fate of an iceberg is determined by a number of processes.
Iceberg trajectories are strongly dependent on the shape and size of the iceberg, so as an iceberg decays, the forces acting on it change. Much of the decay is continuous and takes place in the form of gradual ablation. However, there are other iceberg processes that are inherently less continuous and complicate model representations of iceberg drift and decay. These are \emph{(i)} capsizing, or rolling, and \emph{(ii)} breakup events. This study focuses on the former phenomenon. 

Section 2 reviews previously proposed model representations of iceberg rolling. These rolling criteria assume that an iceberg will roll once a critical width-to-height ratio has been crossed. However, previous studies disagree on the value of this ratio. Section 3 investigates how  rolling impacts the evolution of individual iceberg geometries and their overall lifespan. In Section 4, we consider the effect of rolling in global climate models, focusing on iceberg meltwater distributions at the ocean surface. Conclusions are given in Section 5.

\section{Rolling Criteria}

In what follows, we idealize icebergs to be cuboids, allowing us to consider their stability from a 2D cross-sectional perspective. Icebergs roll along the long axis, $L$, swapping height, $H$, and width, $W$. We further assume that the iceberg is always in isostatic equilibrium, and of uniform ice density, $\ri$. We define the density ratio $\alpha \equiv \ri/\rw$, where $\rw$ is the density of water.

We will show that an iceberg is subject to rolling under infinitesimal perturbations when its width-to-height ratio, $\epsilon \equiv W/H$, is smaller than a given critical value, $\epsilon_c$. For  $\epsilon > \epsilon_c$, on the other hand, the iceberg will return to its original position following an infinitesimal perturbation. 

Here, we derive $\epsilon_c$ by considering the torques acting on an iceberg after it is rotated: for a given clockwise rotation, $\theta$, the center of buoyancy of the submerged part of the iceberg will shift horizontally by a distance $\dx$ relative to the center of gravity (Fig.\ \ref{fig:bergschem}). It can be shown trigonometrically that this is
\be
\dx = \frac{H}{24 \alpha} \left\{\left[12 \alpha \left(\alpha-1\right) + \epsilon^2\right] \sin \theta + \epsilon^2 \sec \theta \tan \theta \right\} \label{eq:dx}
\ee
(see derivation in the Supplemental Information).
Note that equation \eqref{eq:dx} holds only in the range of $\theta$ for which the top surface of the iceberg remains entirely above sea level, $0 \leq \theta < \tan^{-1} \left(2(1-\alpha)/\epsilon\right)$. 

If $\dx > 0$, the counter-clockwise torque that results from the offset of the downward acting gravitational force and the upward acting buoyancy force opposes the direction of rotation, and it acts to restore the original equilibrium. This is the case for shallow wide icebergs (i.e., large $\epsilon$), where the rotation causes additional ice to be submerged to the right of center of gravity (Fig.\ \ref{fig:bergschem}c).  When $\dx < 0$, on the other hand, this torque acts in the direction of rotation, and the iceberg becomes unstable (Fig.\ \ref{fig:bergschem}a). This is the case for tall narrow icebergs (i.e., small $\epsilon$), where the rotation moves submerged ice predominantly to the left of the center of gravity.

We are interested in the response of the iceberg to infinitesimal perturbations, i.e., $\theta \rightarrow 0$. We compute the Maclaurin series of equation \eqref{eq:dx} for small $\theta$, which gives
\be
\dx = \frac{H}{2}\left(\alpha-1+\frac{\epsilon^2}{6\alpha} \right) \theta + O(\theta^2). \label{eq:dx_series}
\ee
The critical width-to-height ratio at which the iceberg becomes unstable under small perturbations can then be defined as $\epsilon_c \equiv \epsilon(\dx = 0)$, for $\theta \rightarrow 0$. From equation \eqref{eq:dx_series} we finally find
\be
\varepsilon_c = \sqrt{6\alpha\left(1 - \alpha \right)}. \label{burt}
\ee

Note that under finite perturbations ($\theta$), icebergs with aspect ratios larger than $\epsilon_c$ may become unstable and capsize \citep{MacAyeal:2003bu, Burton:2012hp}. However, the post-capsize aspect ratio will be less than $ \epsilon_c$, and the iceberg will spontaneously return to its original orientation.


%
An iceberg rolling criterion that is widely used in current iceberg models was introduced by 
\cite{Weeks:1978vi} (henceforth, WM78). WM78 derived a rolling criterion using insights from the ship-building literature, leading to an expression similar to \eqref{burt}. However, WM78 further accounted for an increase in ice density with depth by introducing a correction height, $\Delta$.  In this case, it can be shown that the rolling criterion should become
\be
\epsilon_c = \sqrt{6 \alpha \left(1-\alpha\right) -12\alpha \frac{\Delta}{H}}. \label{eps}
\ee
In agreement with physical intuition, the increase in density with depth leads to a lower value of $\epsilon_c$.
However, we find that the derivation in WM78 erroneously replaced $\Delta$ with $-\Delta$ in the stability criterion [see their equation (9)], an error that does not appear to have been noted in the literature previously and has been carried in many subsequent studies, as discussed below.

WM78 estimated the effective mean density ratio to be $\alpha = 0.81$. By substituting $\alpha = 0.81$ into \eqref{eps}, we find that the sign error in $\Delta$ leads to $\epsilon_c > 1$ for $H<730$ m. Among other issues, this has the unphysical consequence that all icebergs with $H<730$ m will continuously roll (rolling at every model time step) once $\epsilon$ falls below $\epsilon_c$.

\begin{figure}[ht!]
 \begin{center}
 \hspace{-.5 cm} \includegraphics[width=\linewidth]{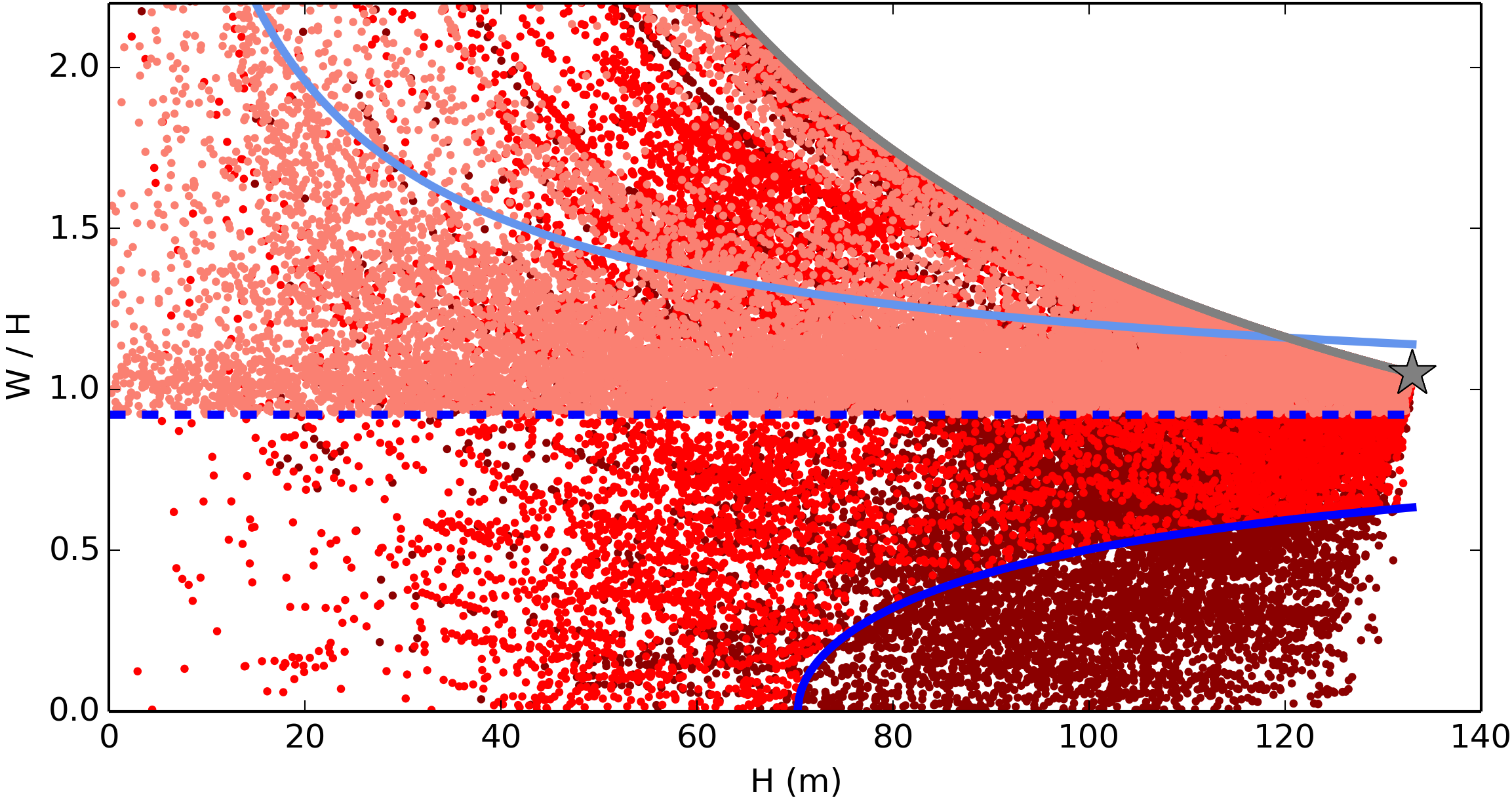}
 \caption{Different rolling criteria, $\epsilon_c(H)$ (blue lines) and simulated iceberg width-to-height ratios (dots). Rolling occurs when $W/H$ for a given iceberg falls below a given blue line. The criterion of WM78 is indicated (light blue), along with a version in which the sign error is corrected (dark blue). Also shown is the criterion of equation \eqref{burt} (dashed blue). Note that any criterion with $\epsilon_c >1$ leads to continuous rolling. Width-to-height ratios are indicated for medium-sized icebergs [size class 3 from \cite{Stern:2016kh}] from GCM model runs with different rolling criteria: non-rolling icebergs (dark red), rolling according to WM78 (red), and rolling according to equation \eqref{burt} (light red). Two years from the simulations are shown, with iceberg dimensions sampled monthly. The original dimensions ($W_0, H_0$) for all icebergs are indicated by the gray star. Icebergs that fall on the gray boundary line experience no side melt and only basal melt (i.e., $W/H = W_0/H$).}
 \label{fig:rollcrit}
 \end{center}
\end{figure}

WM78 were primarily concerned with an idealized iceberg of thickness $H = 200$m, and they approximated the center of gravity correction to be constant, with $\Delta = 6$ m.
It should be emphasized that even the corrected WM78 formulation is not appropriate for use in the continuous evolution of a decaying iceberg. 
First, when the iceberg thickness changes, $\Delta$ should evolve rather than being held constant. This issue is especially egregious for small icebergs, where $\Delta=6$m places the center of gravity of the iceberg so low that the iceberg is unconditionally stable. A consequence of this is that icebergs do not roll when $H<12\Delta/6(1-\alpha) \approx 63$\,m, which allows narrow, pin-like icebergs to occur in the model (Fig.~\ref{fig:rollcrit}).
Second, the density profile of the iceberg assumed in the  WM78 derivation (i.e., that icebergs are densest near the bottom) would be rotated by 90$^\circ$ upon the first instance of rolling, rendering \eqref{eps} no longer appropriate.

However, the criterion of WM78 has subsequently been adopted by many studies of continuous iceberg evolution, which have applied the criterion under a broad range of iceberg dimensions. For example, the rolling criterion (including the sign error and constant value of $\Delta$) is adopted in the seminal study of \cite{Bigg:1997bp}, who consider icebergs of 10 different size classes ranging from 100m $\times$ 66m $\times$ 80m to 1500m $\times$ 1000m $\times$ 360m. In addition to adopting the sign error from WM78, the model of \cite{Bigg:1997bp} also erroneously takes the critical threshold to apply to the ratio $L/H$ rather than $W/H$.
In other words, the horizontal length dimension that is used for the rolling criterion is not the same as the dimension that is rotated from horizontal to vertical when the iceberg rolls.
All of these errors -- the sign error in $\Delta$, the constant value of $\Delta$ associated with $H=200$m  being applied to a wide range of iceberg thicknesses, and the threshold applying to $L/H$ -- are adopted in the more recent iceberg modeling studies of \cite{Gladstone:2001cq}, \cite{Jongma:2009cl}, and \cite{Martin:2010kb}. 
\cite{Martin:2010kb} furthermore erroneously replace $H$ with the draft $\alpha H$ in their representation of the rolling criterion \eqref{eps}. 
The model of \cite{Bigg:1997bp} has been widely adopted, and many recent iceberg modeling studies include these errors in the rolling scheme \citep[e.g.,][]{Death:2006do,Levine:2008jc,Wiersma:2010jj,Jongma:2013hz,Death:2014cv,Roberts:2014ff,vandenBerk:2014cd,Bugelmayer:2015td,Marsh:2015dn,Bigg:2016wo,Merino:2016jm,Stern:2016kh}.


Some studies have also considered the approximation that icebergs have constant density, as in the derivation of equation \eqref{burt}. 
A rolling criterion equivalent to equation \eqref{burt} was derived by \cite{MacAyeal:2003bu}, who evaluated the change in gravitational potential energy under small rotations for an ice-shelf fragment wedged between two segments of an ice shelf. This derivation was later applied to free floating icebergs by  \cite{Burton:2012hp}, who performed laboratory experiments to test the criterion.
In order to take into account the varying density profile of an iceberg, a more complete theory of the density evolution of the iceberg would have to be developed. However, the constant density criterion in equation \eqref{burt} may provide a sufficiently accurate approximation for many purposes. 

Based on observed Larsen A and B ice-shelf densities, \cite{MacAyeal:2003bu} estimated $\epsilon_c \simeq 0.8$, which is close to the value of WM78 and subsequent studies, while \cite{Burton:2012hp} use $\epsilon_c = 0.75$. In the following, we  take $\rho_i = 850$ kg/m$^3$ \citep{Silva:2006wq} 
and $\rho_w = 1025$ kg/m$^3$, such that $\alpha = 0.83$ \citep{Martin:2010kb, Stern:2016kh}, which gives $\epsilon_c \simeq 0.92$. Note that this value of $\rho_i$ is lower than for pure ice due to factors including the snow and firn in the iceberg not being fully compacted.
The stability criterion from equation \eqref{burt}, as well as the original and corrected WM78 schemes, are illustrated in Fig.~\ref{fig:rollcrit}.

 \begin{figure}[ht!]
 \begin{center}
\includegraphics[width=\linewidth]{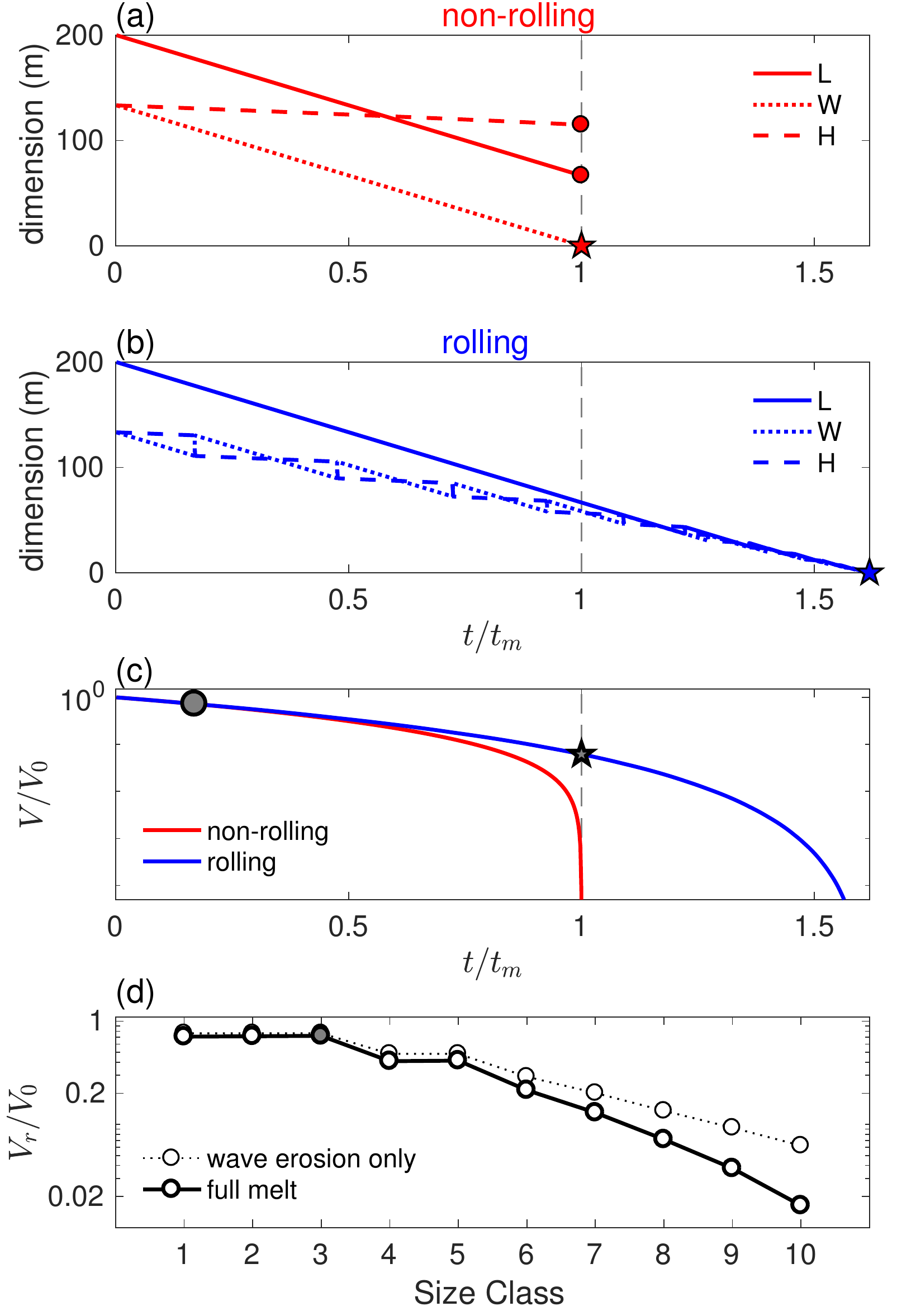}
 \caption{Evolution of iceberg dimensions and volume. (a) Decay of a non-rolling iceberg of size class 3 for fixed forcing values. Shown are dimensions $L$ (solid), $W$ (dotted), and $H$ (dashed) versus time (scaled by time of final melt, $t_m$). The iceberg has fully melted when $W = 0$ (star). Unphysically, $L$ and $H$ have not reached zero (circles).  (b) As in panel a, but including iceberg rolling. Here all dimensions reach zero at the same time, $t_m^*$ (star). (c) Volume scaled by initial volume. Indicated are volume at the time of initial roll (circle), as well as the remaining volume of the rolling iceberg when the non-rolling iceberg has melted (star). (d) Volume (scaled by initial volume) at the time of initial roll for the 10 size classes of Table 1. The dotted line corresponds to the analytic limit of equation (6), which takes $M_v = M_b = 0$. The gray circle corresponds to that in panel c.}
 \label{fig:onset}
 \end{center}
\end{figure}

\section{Impact of Rolling on Individual Icebergs}

In order to study the impact that rolling has on individual icebergs, we will compare two scenarios: \emph{(i)} icebergs undergoing no rolling and \emph{(ii)} icebergs that roll according to the scheme in equation \eqref{burt}. We eschew the WM78 scheme in this section in light of the issues raised above. 
We briefly revisit the WM78 scheme for the GCM simulations in Section 4 in order to estimate the potential bias that this erroneous scheme may have introduced in previous studies. 

\subsection{Iceberg Decay Model}
We first summarize the widely-used decay representation by \cite{Martin:2010kb}, which is based on the earlier work by \cite{Bigg:1997bp}. In this formulation, only three melt processes are considered: \emph{(i)}  wind-driven wave erosion, $M_e$; \emph{(ii)}  turbulent basal melt, $M_b$; and \emph{(iii)}  side wall erosion from buoyant convection, $M_v$. Other processses, such as top and bottom surface melt, are typically small compared to these \citep{Savage:2001hz}. The iceberg dimensions evolve according to $dL/dt = dW/dt = M_e+M_v$ and $dH/dt = M_b$, with iceberg volume given by $V=L \, W \, H$. The individual decay terms are written as follows \citep{Martin:2010kb, Wagner:2016up}:
\begin{eqnarray}
M_e  & = &\tfrac{1}{12}\left(1+ \cos[\pi A_i^3]\right)\left(T_w-T_i\right) S(\va,\vw), \nonumber \\
M_v  & = & b_1 \,T_w + b_2 \, T_w^2, \label{eq:melt} \\
M_b  & = & c \left| \vec{v}_w - \vec{v}_i\right|^{0.8}(T_w-T_i) L^{-0.2},  \nonumber
\end{eqnarray}
where $S$ is the Douglas Sea State, 
$b_1 = 8.8 \times 10^{-8}$ m\,s$^{-1}$$^\circ$C$^{-1}$, $b_2 = 1.5 \times 10^{-8}$ m\,s$^{-1}$$^\circ$C$^{-2}$, $c = 6.7 \times 10^{-6}$ m$^{-2/5}$\,s$^{-1/5}$$^\circ$C$^{-1}$, $T_i$ is the temperature of the ice which is taken to be constant at $-4^\circ$C, $T_w$ is the sea surface temperature (SST),
$A_i$ is the sea-ice concentration, and $\va$, $\vw$, and $\vi$ are the velocities of surface air, surface water, and ice, respectively. 
Note that the expression for $M_e$ includes the wave-dampening influence of sea ice. 

\subsection{Rolling under constant forcing}

In order to estimate how rolling impacts the evolution of the iceberg, it is useful to first consider average forcing conditions. From the GCM simulations of Section 4 we find that average speeds experienced by modern day icebergs are approximately $|\va| = 3$ m/s, $|\vw| = 0.04$ m/s, and $|\vi| = 0.06$ m/s, and average SSTs are approximately $T_w = -1.2^\circ$C. This low temperature is largely due to modern icebergs spending a large fraction of their life surrounded by sea ice (average $A_i = 0.75$).
For these average forcing values, we obtain $M_e \sim 0.3$ m/d, $M_b \sim 0.06$ m/d, and $M_v \sim 0.004$ m/d. Note that this dominance of the wind-driven wave erosion term is in agreement with previous findings \citep[e.g.,][]{Gladstone:2001cq}. The evolution of the individual dimensions of a non-rolling and a rolling iceberg are illustrated in Figs.~\ref{fig:onset}a and b, respectively, using these melt rates and ignoring the effects of iceberg drift. Fig.~\ref{fig:onset} shows that rolling slows down an iceberg's melt: For all realistic values of the density ratio $\alpha$, the critical ratio satisfies $\epsilon_c < 1$, i.e., that $W<H$ at the time of rolling. This implies that the surface area of the sidewalls, $A = 2 H(W+L)$, decreases when the iceberg rolls (because $H$ and $W$ are swapped). As a consequence, the rate of volume loss decreases once rolling begins, since $dV/dt = -A M_e/2$ (Fig.~\ref{fig:onset}c). 

\paragraph{Onset of rolling} As a first step, we set $M_v=M_b=0$, since both terms are considerably smaller than $M_e$. In this approximation, $H$ remains constant. For given initial iceberg dimensions, $H_0$, $W_0$, $L_0$, the critical width at which icebergs roll for the first time is then $W_r = \epsilon_c H_0$. Taking $L_0/W_0 = 3/2$, as is typically done in current iceberg models, we find $L_r = (1/2+\epsilon_c/\epsilon_0)W_0$, where $\epsilon_0 \equiv W_0/H_0$. The volume fraction at first rolling is then 
\be
V_r/V_0 = \tfrac{1}{3}(1+2\epsilon_c/\epsilon_0) (\epsilon_c /\epsilon_0). \label{eq:Vr}
\ee
In order to study the onset of rolling for different initial iceberg sizes, we consider 10 commonly used iceberg size classes, ranging in dimensions from 60 $\times$ 40 $\times$ 40 m to 2200 $\times$ 1467 $\times$ 250 m (Table 1). For size classes 1--3, $\epsilon_0 = 1$ and equation \eqref{eq:Vr} gives $V_r/V_0 = 0.87$, i.e., the iceberg will roll for the first time once it is 13\% decayed. For size classes 4--10, $\epsilon_0$ increases, which leads to a decrease in $V_r$. 
Fig.\,\ref{fig:onset}d summarizes how $V_r$ varies with size class, showing the upper analytical limit \eqref{eq:Vr}, as well as values that take into account nonzero values of $M_v$ and $M_b$. For size class 10, the iceberg begins to roll when it is 98\% decayed (considering all three melt terms).  This highlights that iceberg rolling is most significant for small icebergs.

 \begin{figure}[ht!]
 \begin{center}
\includegraphics[width=\linewidth]{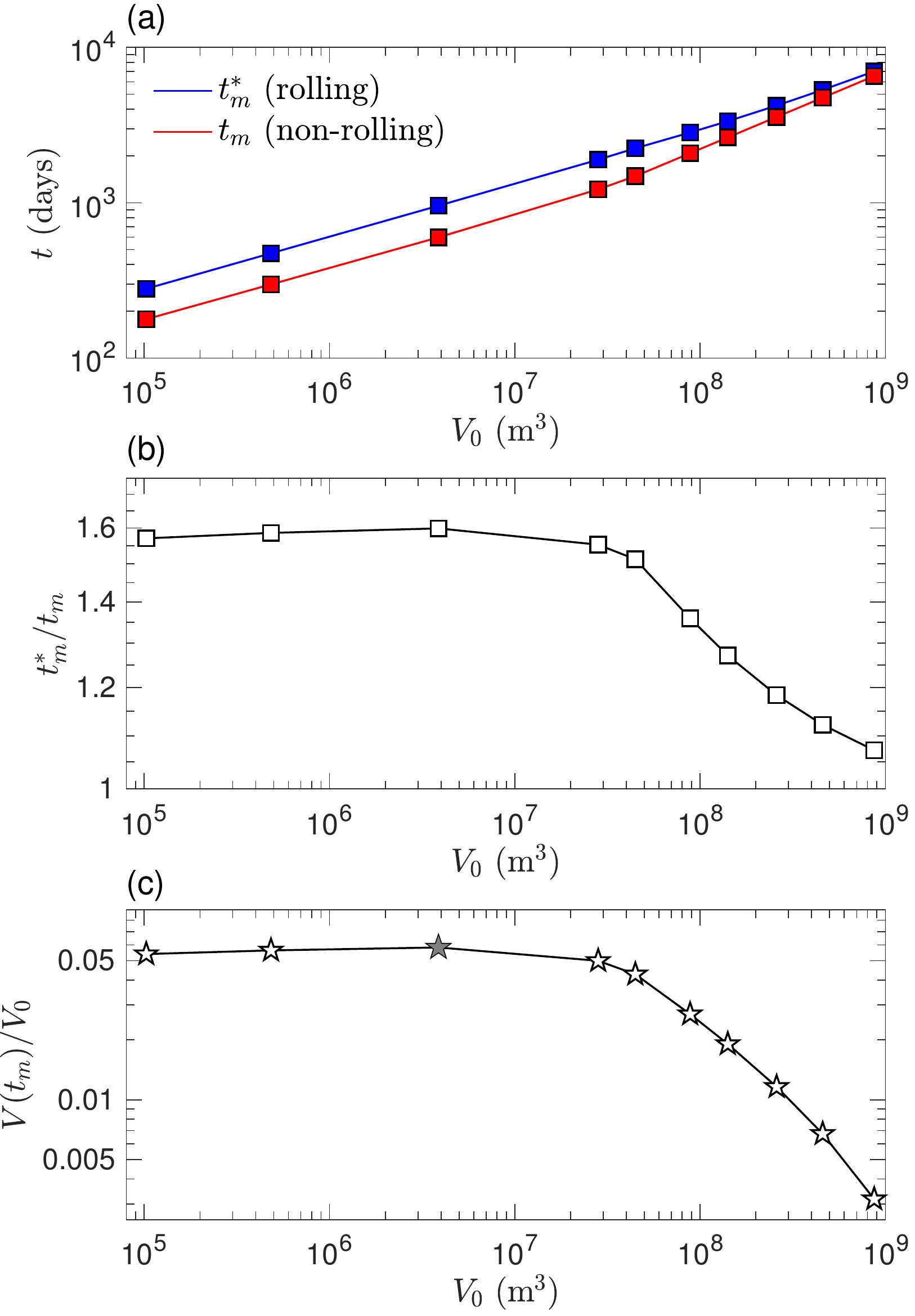}
 \caption{Iceberg life spans.
 (a) Length of iceberg life as a function of initial iceberg size, for size classes 1--10 (see text). (b) Ratio of iceberg lifespan with and without rolling, $t_m^*/t_m$, which shows that rolling icebergs of size classes 1--5 have lifespans about 60\% longer than non-rolling icebergs. For large icebergs, $L$ is sufficiently large that it does not reach zero when $W$ and $H$ have fully melted. This results in the drop off of $t_m^*/t_m$ for large $V_0$. (c) Relative volume of the rolling iceberg that remains at the time when the non-rolling iceberg disappears, $V(t_m)/V_0$. The gray star corresponds to that of Fig.\,3.}
 \label{fig:analytic}
 \end{center}
\end{figure}

\begin{table}
\begin{center}
\begin{tabular}{ccccc} \toprule
    Size Class &  $L_0$  (m) & $W_0 (m)$ &  $H_0$ (m) &  $\epsilon_0$ \\ \midrule
    1  & 60    & 40 & 40   &  1 \\
    2  & 100  & 67 & 67   &  1 \\
    3  & 200  & 133 & 133 & 1 \\  \midrule
    4  & 350  & 233 & 175 & 1.3 \\
    5  & 500  & 333 & 250 & 1.3 \\
    6  & 700  & 467 & 250 & 1.9 \\
    7  & 900  & 600 & 250 & 2.5 \\
    8  & 1200& 800 & 250 & 3.2 \\ 
    9  & 1600& 1067&250 & 4.3 \\
  10  & 2200& 1467&250 & 5.9 \\ \bottomrule
\end{tabular}
\end{center}
\vspace{-.2 cm} \caption{Initial iceberg dimensions and $\epsilon_0 \equiv W_0/H_0$ for the 10  size classes used here (adapted from \cite{Stern:2016kh}, which is based on \cite{Gladstone:2001cq}).}
\end{table}

\paragraph{Impact of rolling on iceberg life span}
Under the approximation that $M_b = M_v = 0$, a \emph{non-rolling} iceberg would completely melt at time $t_m = W_0/M_e$, regardless of $L_0$. 
The time of complete melt for a \emph{rolling} iceberg, on the other hand, can be readily shown to be
\be
t_m^* = [5/4 +1/(2\epsilon_0)](W_0/M_e), \label{eq:tm}
\ee
where we again have fixed $L_0/W_0 = 3/2$. The relative life span of rolling versus non-rolling icebergs is then $t_m^*/t_m = 5/4 +1/2\epsilon_0 > 1$, i.e., rolling icebergs always live longer (Fig.~\ref{fig:analytic}a). Hence icebergs with a large initial width-to-height ratio, such that $\epsilon_0 \rightarrow \infty$, live 25\% longer. 
For size classes 1--3 icebergs would live 75\% longer due to rolling. Fig.~\ref{fig:analytic}b shows that, by accounting for $M_b$ and $M_v$, the actual increase in life span due to rolling is closer to 60\%. Note further that equation \eqref{eq:tm} assumes constant forcing and neglects the impact of rolling on the drift dynamics, which will also impact melt rates (see Section 4). Fig.~\ref{fig:analytic}b shows that $t_m^*/t_m$ decreases rapidly for large icebergs. In fact, $t_m^*/t_m$ falls below the limit 5/4, which is due to $W$ and $H$ reaching zero before $L$ has melted completely. This highlights an unphysical feature in many current iceberg models: the models ultimately end up with long, infinitely thin iceberg slithers. 
Finally, even though rolling icebergs can live substantially longer, the remaining volume at time $t_m$ (when the non-rolling iceberg has completely melted) is small, with $V(t_m)/V_0 < 6\%$ (Fig.~\ref{fig:analytic}c).

\section{Impact of Rolling in Iceberg--Climate Model Simulations}

In what follows we compare the meltwater release in model simulations 
with rolling and non-rolling icebergs, using \emph{(1)} an idealized iceberg drift and decay model and \emph{(2)} a comprehensive coupled GCM. 

\subsection{Idealized offline iceberg model}

 \begin{figure*}[!t]
 \begin{center}
 \includegraphics[width=0.8\linewidth]{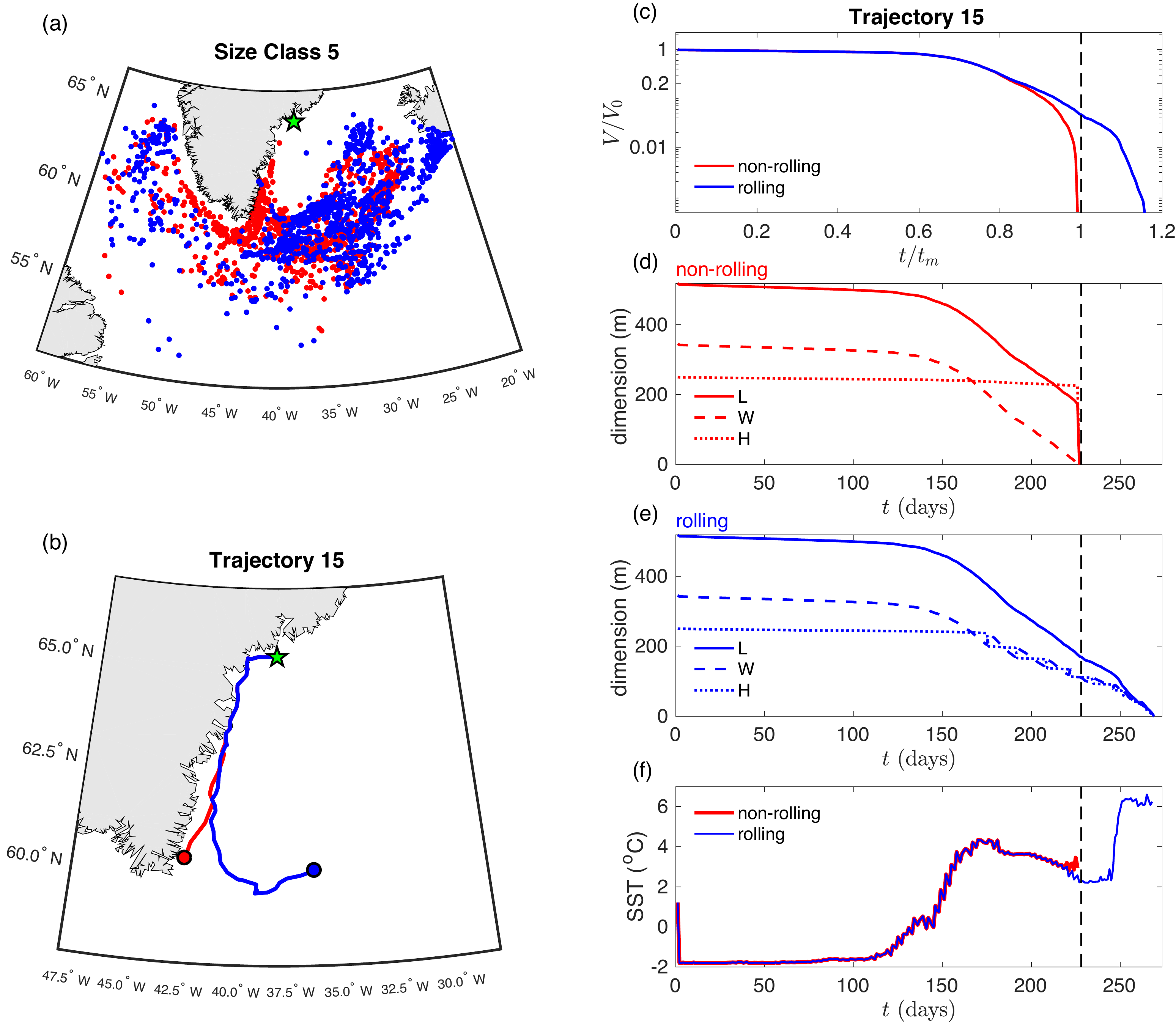}
 \caption{Results from the idealized, offline iceberg simulations of Section 3a. (a) Locations of final melt for size class 5 rolling icebergs (blue) and non-rolling icebergs (red). All icebergs are released in the region marked by the green star (near the mouth of Sermilik Fjord, East Greenland). (b) Detail of a pair of rolling and non-rolling iceberg trajectories (trajectory 15 in the simulations). (c) Evolution of volume (scaled by initial volume, $V(t)/V_0$, corresponding to the trajectories of (b). The time of complete melt of the non-rolling iceberg, $t = t_m$, is indicated (dashed black). (d) Evolution of iceberg dimensions for the non-rolling trajectory 15, and (e) for the rolling trajectory 15. (f) SST conditions experienced by the two icebergs.}
 \label{fig:WE_SC5}
 \end{center}
\end{figure*}

We use a recently developed drift model which evolves iceberg velocity under the influence of air drag, water drag, the pressure gradient force, and the Coriolis force \citep{Wagner:2016up}. This formulation is somewhat idealized compared to previous iceberg models \citep[e.g.,][]{Bigg:1997bp,Gladstone:2001cq,Roberts:2014ff}, allowing an analytical solution for iceberg velocity as a function of surface air and water velocities.  The model operates in an offline mode, meaning that the iceberg trajectories are computed as non-interactive Lagrangian particles, using precomputed input surface velocity and SST fields. This allows for a rapid integration of large numbers of iceberg trajectories. The drift model is coupled to the decay model of equation \eqref{eq:melt}. 

The precomputed input fields are taken from NASA's Estimating the Circulation and Climate of the Ocean Phase II (ECCO2) global ocean state estimate
\citep{Menemenlis:2008ve}. We perform two sets of simulations, one with iceberg rolling, using the  scheme in equation \eqref{burt}, and one without iceberg rolling. In order to avoid canceling effects due to different iceberg release locations (see Section 4b), we initialize icebergs in a small region just off the coast of Sermilik Fjord, East Greeland, into which Helheim Glacier, one of Greenland's largest outlet glaciers, drains (green star in Fig.~\ref{fig:WE_SC5}a). The release location of each iceberg is set to be at the center of a grid box which is randomly chosen from a $5 \times 4$ region of grid boxes. Since this region only has substantial sea ice for a brief period of the year, we set $A_i = 0$ in the following simulations.

We release 5000 icebergs of each of the 10 iceberg size classes specified in \cite{Stern:2016kh}, at a rate of 14/day over the input year 1992. Icebergs are released at identical locations and times in the two simulations, and they are tracked until fully decayed. 



Fig.~\ref{fig:WE_SC5} illustrates the example of size class 5, which is broadly representative of the other size classes (not shown). The longer life span and farther reach of rolling icebergs is readily seen in these simulations. The blue and red dots in Fig.~\ref{fig:WE_SC5}a show the final melting points of rolling and non-rolling icebergs, respectively.  The non-rolling icebergs are found to typically melt closer to the release location, compared to the rolling icebergs. Figs.~\ref{fig:WE_SC5}b--f illustrate the evolution of a representative pair of rolling and non-rolling trajectories. The rolling iceberg survives $\sim$20\% longer (Fig.~\ref{fig:WE_SC5}c). The slow rate of decay over the first half of the icebergs' lives (Fig.~\ref{fig:WE_SC5}c--e) is due to cold winter conditions, and the melt rates speed up substantially once summer arrives, as can also be seen from the rise in temperatures (Fig.~\ref{fig:WE_SC5}f). Figs.~\ref{fig:WE_SC5}d, e illustrate that the evolution of the simulated iceberg dimensions is comparable to that of the analytic version (Fig.~\ref{fig:onset}a, b). The shorter relative life span of the rolling iceberg ($t_m^*/t_m = 1.2$), compared to the analytic value ($t_m^*/t_m \simeq 1.5$ in Fig 4b), is likely due to a number of factors, including higher values of $M_e$ once the iceberg reaches the open ocean and warmer temperatures (Fig.~\ref{fig:WE_SC5}f). We note that the rolling also has an immediate effect on the iceberg's drift direction. This is due to rolling changing the aspect ratio of the iceberg, and hence changing the momentum balance. This explains why the iceberg trajectories diverge upon the first rolling event ($t = 161$ d), long before the final melt of the non-rolling iceberg ($t=228$ d, Fig.~\ref{fig:WE_SC5}b). Furthermore, the change in aspect ratio from rolling causes the iceberg drift to slow (not shown), which also mitigates the farther spread of the rolling icebergs.


In order to assess the potential large-scale and long-term impacts of rolling, we will consider simulations with a coupled GCM in the next section. This will allow us to address the question whether the biases of Fig.~\ref{fig:WE_SC5}a are apparent in comprehensive model runs, or whether they are small compared to the variability (``noise") of the system.

\subsection{GCM Simulations}

 \begin{figure*}[!t]
 \begin{center}
 \includegraphics[width=0.7\linewidth]{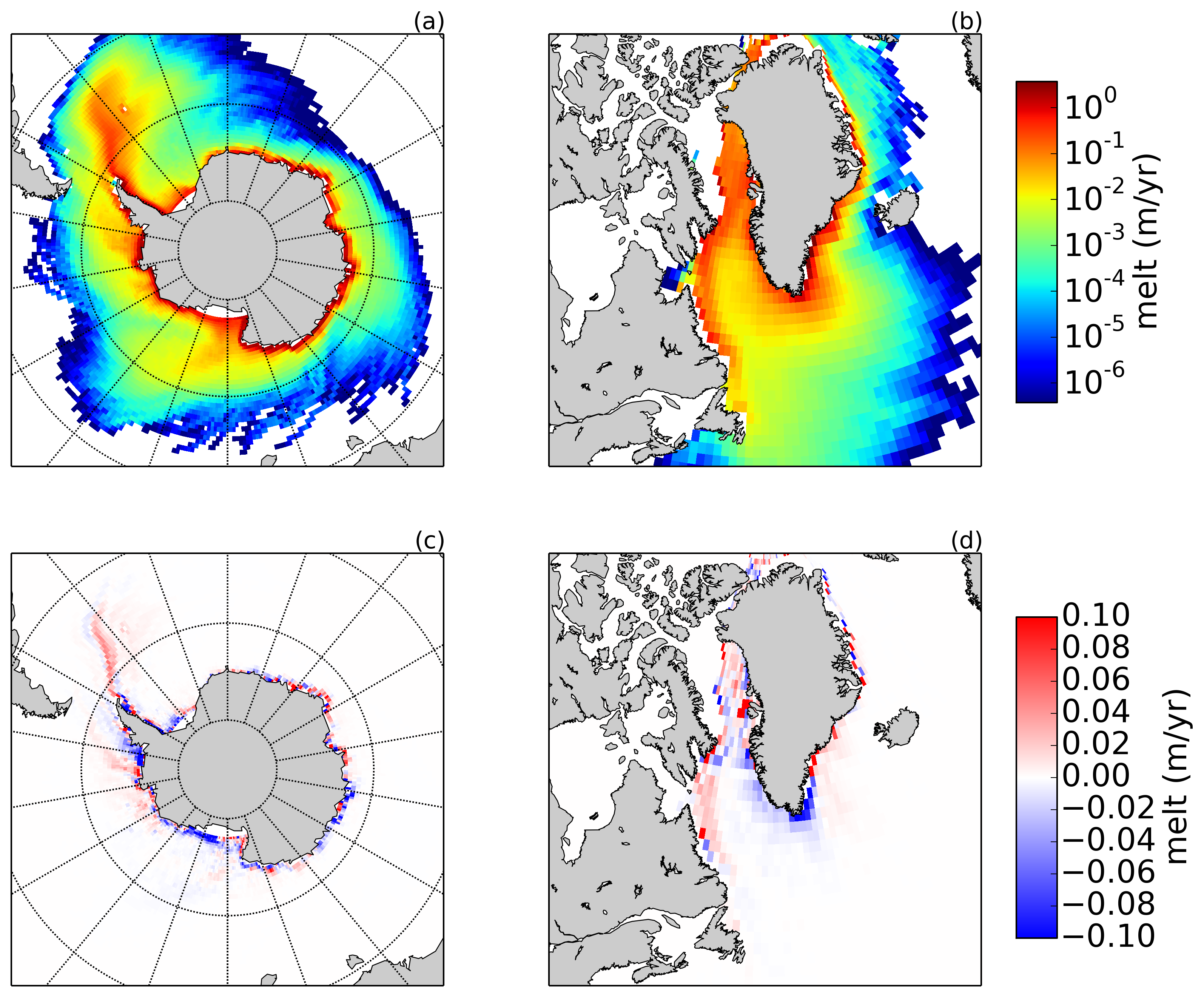}
 \caption{GCM freshwater fluxes. (a) Antarctic freshwater flux from iceberg melt, averaged over 100 years, using the iceberg rolling scheme of equation \eqref{burt}. (b) As in (a) but for Greenland. Note the logarithmic scale in both panels. (c) and (d) Differences in freshwater flux, calculated from the simulations with iceberg rolling (a,b) minus those without rolling (not shown). No regions show statistically significant differences. All fluxes are given in meters per year.}
 \label{fig:GCM_flux}
 \end{center}
\end{figure*}

We use the Geophysical Fluid Dynamics Laboratory (GFDL) coupled climate model CM2G \citep{Delworth:2006dj,Dunne:2012bo}, which includes the following components: AM2 atmosphere model, MOM6 ocean model, SIS2 sea-ice model, and LM3 land model, as well as the iceberg component detailed in \cite{Martin:2010kb}. The ocean model uses a $1^{\circ} \times 1^{\circ}$ horizontal grid, with 63 isopycnal layers in the vertical, and the atmospheric model has a $2^{\circ}$ horizontal resolution. The model setup is as described in \cite{Stern:2016kh}, except for the iceberg rolling scheme. The icebergs in these simulations are fully coupled to the climate model, so that the melt water from the icebergs freshens and cools the ocean surface, which can lead to feedback effects acting on the greater climate system.

In all of the GCM simulations icebergs are released into the global ocean according to the scheme of \cite{Stern:2016kh}, using the same 10 iceberg size classes, size class distributions, and release locations. These locations encompass most major ice shelves and outlet glaciers in the Arctic and Antarctic. The simulations are run for 150 years each. 

We perform three simulations: one with iceberg rolling using the scheme in equation \eqref{burt}, one with rolling using the WM78 scheme but with the sign error of $\Delta$ corrected, and one with no iceberg rolling. Icebergs in these simulations fill the phase space of available aspect ratios ($W/H$), bound by the critical rolling ratio $\epsilon_c$ and the limits $W_{0}/H$ and $W/H_0$ (see Fig.~2). The latter two limits indicate pure basal melt and pure sidewall melt, respectively. The large spread of simulated ratios indicates the wide range of ocean conditions and melt rates experienced by icebergs in different parts of the globe and at different times of the year.  

The simulations with equation \eqref{burt}, which have much earlier rolling than in the corrected WM78 scheme, can be regarded as an upper bound of how much impact rolling may have, and we will in the following focus on comparing simulations with non-rolling icebergs and icebergs that roll according to equation \eqref{burt}.

Figs.~\ref{fig:GCM_flux}a and b show the freshwater flux due to iceberg melting around Antarctica and Greenland, averaged over the final 100 years of the simulations with the scheme from equation \eqref{burt}. These distributions are qualitatively similar to those in previous studies \citep[e.g.,][which used the original WM78 scheme]{Martin:2010kb,Stern:2016kh}, which indicates that the differences in rolling schemes do not strongly affect the large-scale freshwater flux distribution. This is further supported when we look at the difference in freshwater flux between the present rolling and no-rolling simulations (Fig.~\ref{fig:GCM_flux}c, d). While we do observe small differences, particularly in coastal regions, none of these differences are statistically significant at the 95\% confidence level over the time period of integration. The same result is found for simulations with the corrected WM78 scheme (not shown). The relatively small rolling signal appears to be due to \emph{(i)} the high natural variability of the overall freshwater forcing (which is, for example,  strongly influenced by sea ice which itself is highly variable), \emph{(ii)} the fact that rolling affects icebergs only after much of their initial volume has already melted, as discussed above (Fig. 3). 

\begin{figure}[!t]
 \begin{center}
 \includegraphics[width=\linewidth]{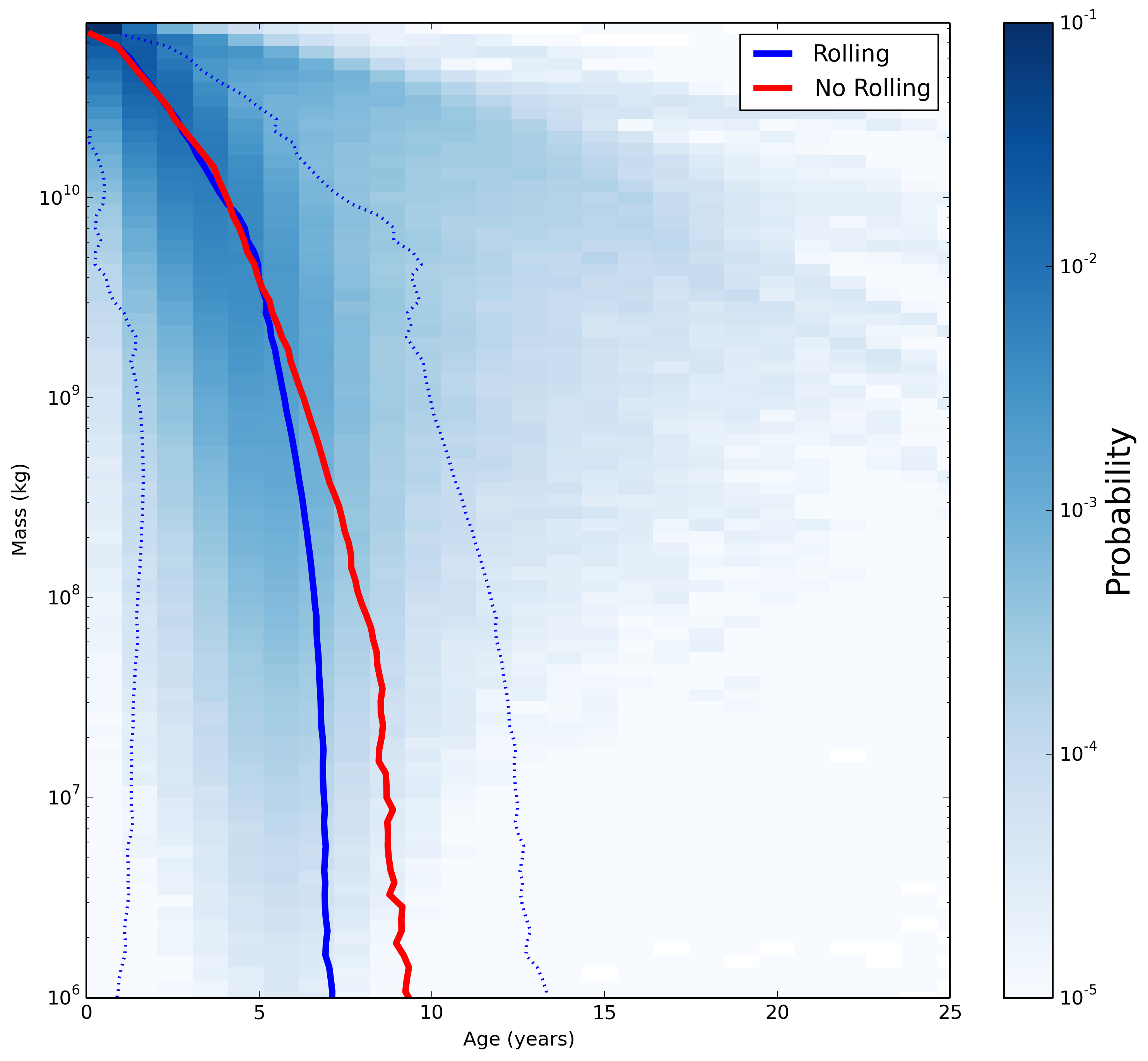}
 \caption{Evolution of iceberg mass from GCM simulations for size class 10 (considering all icebergs in the global simulations). Shading shows relative probabilities of iceberg mass as a function of iceberg age for rolling icebergs [using the rolling scheme of equation \eqref{burt}]. Further indicated are the mean rolling (blue line) and no-rolling (red line) iceberg masses over time, and the spread of the rolling icebergs (one standard deviation, dotted blue).}
 \label{fig:GCMage}
 \end{center}
\end{figure}

The small differences in overall freshwater flux do not rule out that rolling icebergs may on average live longer or travel farther than their non-rolling counterparts, as may be expected from Sections 2 and 3. However, the wide range of ocean conditions which icebergs experience, together with the large variability of the fully-coupled climate model, give rise to a large spread of iceberg lifetimes (Fig.~\ref{fig:GCMage}). Modeled iceberg lifetimes are, for example, sensitive to sea ice conditions, and can be greatly increased by icebergs becoming stuck along the Antarctic and Greenlandic coasts (factors that are not taken into account in the idealized model of Section 4a).
In fact, we find that in the GCM non-rolling icebergs live on average slightly \emph{longer} than their rolling counterparts (contrary to our earlier findings), although Fig.~\ref{fig:GCMage} shows that this difference is small compared to the spread of lifetimes. This slightly counter-intuitive result may be due to several factors: First, the large natural variability and long time scales of the fully-coupled system result in different sea ice conditions for the two simulations, especially in the northern hemisphere (not shown).  While these differences are not statistically significant when averaged over the 150 year simulation period, we do overall observe slightly warmer surface temperatures (and decreased sea ice) in the rolling simulation, which leads to more rapidly decaying icebergs. In addition to natural variability, iceberg lifetimes are influenced by their trajectories, as discussed above. After rolling, icebergs tend to spread more readily away from the coast (both in Antarctica and Greenland, see also Fig.~5a), causing them to be exposed to warmer waters. The combination of warmer conditions in the rolling simulation and more offshore trajectories for rolling icebergs appear to be enough to offset the decreases in iceberg decay caused by the geometric considerations of Section 3.

In summary, the findings of this section suggest that the large-scale biases that are introduced in coupled climate simulations by the use of inadequate iceberg rolling schemes are relatively small. 


\section{Conclusion} \label{sec:conc}

In this study, we have addressed some of the details of how to best account for the rolling of icebergs in models that explicitly represent icebergs. We have shown that a widely used rollover criterion, based on  work by \cite{Weeks:1978vi}, erroneously describes the rolling of icebergs for a broad range of iceberg sizes. For large-scale studies, the results presented here suggest that this criterion should be replaced by the more physical scheme in equation \eqref{burt}. For studies concerned with individual iceberg trajectories, a more sophisticated rolling scheme may be required. 

We have further shown that iceberg rolling has an overall small impact on the large-scale iceberg meltwater flux.
Nevertheless, we find that rolling strongly impacts the drift and decay of individual icebergs, in particular relatively small icebergs (length $\lesssim 500$ m). For example, we find that for fixed surface velocities and SSTs, rolling icebergs typically live substantially longer and drift farther than non-rolling icebergs. On the one hand, this is highly relevant for operational forecast models of iceberg drift. On the other hand, it may substantially impact the results of high-resolution regional climate model simulations, which are receiving increased attention, and which are likely to use iceberg rollover representations such as the schemes discussed here. 

\section*{Acknowledgments}
We are grateful to Jeff Severinghaus and Ralph Keeling for helpful discussions during the development of this work.
This work was supported by National Science Foundation grant OCE-1357078.

\section*{References}
\bibliographystyle{elsarticle-harv} 
\bibliography{Rolling}

\begin{thebibliography}{39}
\expandafter\ifx\csname natexlab\endcsname\relax\def\natexlab#1{#1}\fi
\expandafter\ifx\csname url\endcsname\relax
  \def\url#1{\texttt{#1}}\fi
\expandafter\ifx\csname urlprefix\endcsname\relax\def\urlprefix{URL }\fi

\bibitem[{Bigg(2016)}]{Bigg:2016wo}
Bigg, G., Jan. 2016. {Icebergs: Their Science and Links to Global Change}.
  Cambridge University Press.

\bibitem[{Bigg et~al.(1997)Bigg, Wadley, Stevens, and Johnson}]{Bigg:1997bp}
Bigg, G.~R., Wadley, M.~R., Stevens, D.~P., Johnson, J.~A., Oct. 1997.
  {Modelling the dynamics and thermodynamics of icebergs}. Cold Regions Science
  and Technology 26~(2), 113--135.

\bibitem[{Broecker(1994)}]{Broecker:1994em}
Broecker, W.~S., Dec. 1994. {Massive iceberg discharges as triggers for global
  climate change}. Nature 372~(6505), 421--424.

\bibitem[{B{\"u}gelmayer et~al.(2015)B{\"u}gelmayer, Roche, and
  Renssen}]{Bugelmayer:2015td}
B{\"u}gelmayer, M., Roche, D.~M., Renssen, H., 2015. {How do icebergs affect
  the Greenland ice sheet under pre-industrial conditions? - a model study with
  a fully coupled ice-sheet-climate model}. Cryosphere 9~(3), 821--835.

\bibitem[{Burton et~al.(2012)Burton, Amundson, Abbot, Boghosian, Cathles,
  Correa-Legisos, Darnell, Guttenberg, Holland, and MacAyeal}]{Burton:2012hp}
Burton, J.~C., Amundson, J.~M., Abbot, D.~S., Boghosian, A., Cathles, L.~M.,
  Correa-Legisos, S., Darnell, K.~N., Guttenberg, N., Holland, D.~M., MacAyeal,
  D.~R., 2012. {Laboratory investigations of iceberg capsize dynamics, energy
  dissipation and tsunamigenesis}. Journal of Geophysical Research 117, F01007.

\bibitem[{Cooke et~al.(2012)Cooke, Dunne, Harrison, Malyshev, Milly, Sentman,
  Samuels, Spelman, Winton, Dunne, John, Adcroft, Griffies, Hallberg,
  Shevliakova, Stouffer, Krasting, Phillipps, Wittenberg, and
  Zadeh}]{Dunne:2012bo}
Cooke, W., Dunne, K.~A., Harrison, M.~J., Malyshev, S.~L., Milly, P. C.~D.,
  Sentman, L.~T., Samuels, B.~L., Spelman, M.~J., Winton, M., Dunne, J.~P.,
  John, J.~G., Adcroft, A.~J., Griffies, S.~M., Hallberg, R.~W., Shevliakova,
  E., Stouffer, R.~J., Krasting, J.~P., Phillipps, P.~J., Wittenberg, A.~T.,
  Zadeh, N., Apr. 2012. {GFDL{\textquoteright}s ESM2 Global Coupled
  Climate{\textendash}Carbon Earth System Models. Part I: Physical Formulation
  and Baseline Simulation Characteristics}. dx.doi.org 25~(19), 6646--6665.

\bibitem[{Copland et~al.(2007)Copland, Mueller, and Weir}]{Copland:2007du}
Copland, L., Mueller, D.~R., Weir, L., 2007. {Rapid loss of the ayles ice
  shelf, ellesmere island, Canada}. Geophysical Research Letters 34~(21),
  L21501.

\bibitem[{Death et~al.(2006)Death, Siegert, Bigg, and Wadley}]{Death:2006do}
Death, R., Siegert, M.~J., Bigg, G.~R., Wadley, M.~R., 2006. {Modelling iceberg
  trajectories, sedimentation rates and meltwater input to the ocean from the
  Eurasian Ice Sheet at the Last Glacial Maximum}. Palaeogeography,
  Palaeoclimatology, Palaeoecology 236~(1-2), 135--150.

\bibitem[{Death et~al.(2014)Death, Wadham, Monteiro, Le~Brocq, Tranter,
  Ridgwell, Dutkiewicz, and Raiswell}]{Death:2014cv}
Death, R., Wadham, J.~L., Monteiro, F., Le~Brocq, A.~M., Tranter, M., Ridgwell,
  A., Dutkiewicz, S., Raiswell, R., 2014. {Antarctic ice sheet fertilises the
  Southern Ocean}. Biogeosciences 11~(10), 2635--2643.

\bibitem[{Delworth et~al.(2006)Delworth, Broccoli, Rosati, Stouffer, Balaji,
  Beesley, Cooke, Dixon, Dunne, Dunne, Durachta, Findell, Ginoux, Gnanadesikan,
  Gordon, Griffies, Gudgel, Harrison, Held, Hemler, Horowitz, Klein, Knutson,
  Kushner, Langenhorst, Lee, Lin, Lu, Malyshev, Milly, Ramaswamy, Russell,
  Schwarzkopf, Shevliakova, Sirutis, Spelman, Stern, Winton, Wittenberg, Wyman,
  Zeng, and Zhang}]{Delworth:2006dj}
Delworth, T.~L., Broccoli, A.~J., Rosati, A., Stouffer, R.~J., Balaji, V.,
  Beesley, J.~A., Cooke, W.~F., Dixon, K.~W., Dunne, J., Dunne, K.~A.,
  Durachta, J.~W., Findell, K.~L., Ginoux, P., Gnanadesikan, A., Gordon, C.~T.,
  Griffies, S.~M., Gudgel, R., Harrison, M.~J., Held, I.~M., Hemler, R.~S.,
  Horowitz, L.~W., Klein, S.~A., Knutson, T.~R., Kushner, P.~J., Langenhorst,
  A.~R., Lee, H.-C., Lin, S.-J., Lu, J., Malyshev, S.~L., Milly, P. C.~D.,
  Ramaswamy, V., Russell, J., Schwarzkopf, M.~D., Shevliakova, E., Sirutis,
  J.~J., Spelman, M.~J., Stern, W.~F., Winton, M., Wittenberg, A.~T., Wyman,
  B., Zeng, F., Zhang, R., Mar. 2006. {GFDL's CM2 Global Coupled Climate
  Models. Part I: Formulation and Simulation Characteristics}. Journal of
  Climate 19~(5), 643--674.

\bibitem[{Duprat et~al.(2016)Duprat, Bigg, and Wilton}]{Duprat:2016hw}
Duprat, L. P. A.~M., Bigg, G.~R., Wilton, D.~J., Mar. 2016. {Enhanced Southern
  Ocean marine productivity due to fertilization by giant icebergs}. Nature
  Geoscience 9~(3), 219--221.

\bibitem[{Gladstone et~al.(2001)Gladstone, Bigg, and
  Nicholls}]{Gladstone:2001cq}
Gladstone, R.~M., Bigg, G.~R., Nicholls, K.~W., 2001. {Iceberg trajectory
  modeling and meltwater injection in the Southern Ocean}. Journal of
  Geophysical Research 106~(C9), 19903--19915.

\bibitem[{Henderson and Loe(2016)}]{Henderson:2016wa}
Henderson, J., Loe, J. S.~P., May 2016. {The Prospects and Challenges for
  Arctic Oil Development}. Oil, Gas {\&} Energy Law Journal (OGEL) 14~(2).

\bibitem[{Hunke and Comeau(2011)}]{Hunke:2011fx}
Hunke, E.~C., Comeau, D., May 2011. {Sea ice and iceberg dynamic interaction}.
  Journal of Geophysical Research: Atmospheres 116~(C5), C05008.

\bibitem[{Jongma et~al.(2009)Jongma, Driesschaert, Fichefet, Goosse, and
  Renssen}]{Jongma:2009cl}
Jongma, J.~I., Driesschaert, E., Fichefet, T., Goosse, H., Renssen, H., 2009.
  {The effect of dynamic-thermodynamic icebergs on the Southern Ocean climate
  in a three-dimensional model}. Ocean Modelling 26~(1-2), 104--113.

\bibitem[{Jongma et~al.(2013)Jongma, Renssen, and Roche}]{Jongma:2013hz}
Jongma, J.~I., Renssen, H., Roche, D.~M., Mar. 2013. {Simulating Heinrich event
  1 with interactive icebergs}. Climate Dynamics 40~(5-6), 1373--1385.

\bibitem[{Joughin et~al.(2014)Joughin, Smith, and Medley}]{Joughin:2014ew}
Joughin, I., Smith, B.~E., Medley, B., 2014. {Marine Ice Sheet Collapse
  Potentially Under Way for the Thwaites Glacier Basin, West Antarctica}.
  Science 344~(6185), 735--738.

\bibitem[{Levine and Bigg(2008)}]{Levine:2008jc}
Levine, R.~C., Bigg, G.~R., 2008. {Sensitivity of the glacial ocean to Heinrich
  events from different iceberg sources, as modeled by a coupled
  atmosphere-iceberg-ocean model}. Paleoceanography 23~(4), --n/a.

\bibitem[{MacAyeal et~al.(2003)MacAyeal, Scambos, Hulbe, and
  Fahnestock}]{MacAyeal:2003bu}
MacAyeal, D.~R., Scambos, T.~A., Hulbe, C.~L., Fahnestock, M.~A., 2003.
  {Catastrophic ice-shelf break-up by an ice-shelf-fragment-capsize mechanism}.
  Journal of Glaciology 49~(164), 22--36.

\bibitem[{Marsh et~al.(2015)Marsh, Ivchenko, Skliris, Alderson, Bigg, Madec,
  Blaker, Aksenov, Sinha, Coward, Le~Sommer, Merino, and
  Zalesny}]{Marsh:2015dn}
Marsh, R., Ivchenko, V.~O., Skliris, N., Alderson, S., Bigg, G.~R., Madec, G.,
  Blaker, A.~T., Aksenov, Y., Sinha, B., Coward, A.~C., Le~Sommer, J., Merino,
  N., Zalesny, V.~B., 2015. {NEMO-ICB (v1.0): interactive icebergs in the NEMO
  ocean model globally configured at eddy-permitting resolution}. Geoscientific
  Model Development 8~(5), 1547--1562.

\bibitem[{Martin and Adcroft(2010)}]{Martin:2010kb}
Martin, T., Adcroft, A., 2010. {Parameterizing the fresh-water flux from land
  ice to ocean with interactive icebergs in a coupled climate model}. Ocean
  Modelling 34~(3-4), 111--124.

\bibitem[{Menemenlis et~al.(2008)Menemenlis, Campin, and
  Heimbach}]{Menemenlis:2008ve}
Menemenlis, D., Campin, J.~M., Heimbach, P., 2008. {ECCO2: High resolution
  global ocean and sea ice data synthesis}. Mercator Ocean Quarterly Newsletter
  31, 13--21.

\bibitem[{Merino et~al.(2016)Merino, Le~Sommer, Durand, Jourdain, Madec,
  Mathiot, and Tournadre}]{Merino:2016jm}
Merino, N., Le~Sommer, J., Durand, G., Jourdain, N.~C., Madec, G., Mathiot, P.,
  Tournadre, J., Aug. 2016. {Antarctic icebergs melt over the Southern Ocean:
  Climatology and impact on sea ice}. Ocean Modelling 104, 99--110.

\bibitem[{Pizzolato et~al.(2014)Pizzolato, Howell, Derksen, Dawson, and
  Copland}]{Pizzolato:2014cy}
Pizzolato, L., Howell, S. E.~L., Derksen, C., Dawson, J., Copland, L., Mar.
  2014. {Changing sea ice conditions and marine transportation activity in
  Canadian Arctic waters between 1990 and 2012}. Climatic Change 123~(2),
  161--173.

\bibitem[{Rignot and Kanagaratnam(2006)}]{Rignot:2006fm}
Rignot, E., Kanagaratnam, P., Feb. 2006. {Changes in the Velocity Structure of
  the Greenland Ice Sheet}. Science 311~(5763), 986--990.

\bibitem[{Rignot et~al.(2011)Rignot, Velicogna, van~den Broeke, Monaghan, and
  Lenaerts}]{Rignot:2011hi}
Rignot, E., Velicogna, I., van~den Broeke, M.~R., Monaghan, A., Lenaerts, J.,
  2011. {Acceleration of the contribution of the Greenland and Antarctic ice
  sheets to sea level rise}. Geophysical Research Letters 38, L05503.

\bibitem[{Roberts et~al.(2014)Roberts, Valdes, and Payne}]{Roberts:2014ff}
Roberts, W. H.~G., Valdes, P.~J., Payne, A.~J., 2014. {A new constraint on the
  size of Heinrich Events from an iceberg/sediment model}. Earth and Planetary
  Science Letters 386, 1--9.

\bibitem[{Savage(2001)}]{Savage:2001hz}
Savage, S.~B., Dec. 2001. {Aspects of Iceberg Deterioration and Drift}. In:
  Geomorphological Fluid Mechanics. Springer Berlin Heidelberg, Berlin,
  Heidelberg, pp. 279--318.

\bibitem[{Silva et~al.(2006)Silva, Bigg, and Nicholls}]{Silva:2006wq}
Silva, T., Bigg, G.~R., Nicholls, K.~W., 2006. {Contribution of giant icebergs
  to the Southern Ocean freshwater flux}. Journal of Geophysical Research
  111~(C3).

\bibitem[{Smith et~al.(2013)Smith, Sherman, Shaw, and Sprintall}]{Smith:2013cu}
Smith, K. L.~J., Sherman, A.~D., Shaw, T.~J., Sprintall, J., 2013. {Icebergs as
  Unique Lagrangian Ecosystems in Polar Seas}. Annual Review of Marine Science
  5~(1), 269--287.

\bibitem[{Stern et~al.(2016)Stern, Adcroft, and Sergienko}]{Stern:2016kh}
Stern, A.~A., Adcroft, A., Sergienko, O., Jul. 2016. {The effects of Antarctic
  iceberg calving-size distribution in a global climate model}. Journal of
  Geophysical Research: Oceans 121, JC011835.

\bibitem[{Stern et~al.(2015)Stern, Johnson, Holland, Wagner, Wadhams, Bates,
  Abrahamsen, Nicholls, Crawford, Gagnon, and Tremblay}]{Stern:2015bo}
Stern, A.~A., Johnson, E., Holland, D.~M., Wagner, T. J.~W., Wadhams, P.,
  Bates, R., Abrahamsen, E.~P., Nicholls, K.~W., Crawford, A., Gagnon, J.,
  Tremblay, J.-E., Aug. 2015. {Wind-driven upwelling around grounded tabular
  icebergs}. Journal of Geophysical Research: Oceans 120~(8), 5820--5835.

\bibitem[{Stokes et~al.(2015)Stokes, Tarasov, Blomdin, Cronin, Fisher,
  Gyllencreutz, Hattestrand, Heyman, Hindmarsh, Hughes, Jakobsson, Kirchner,
  Livingstone, Margold, Murton, Noormets, Peltier, Peteet, Piper, Preusser,
  Renssen, Roberts, Roche, Saint-Ange, Stroeven, and Teller}]{Stokes:2015dt}
Stokes, C.~R., Tarasov, L., Blomdin, R., Cronin, T.~M., Fisher, T.~G.,
  Gyllencreutz, R., Hattestrand, C., Heyman, J., Hindmarsh, R. C.~A., Hughes,
  A. L.~C., Jakobsson, M., Kirchner, N., Livingstone, S.~J., Margold, M.,
  Murton, J.~B., Noormets, R., Peltier, W.~R., Peteet, D.~M., Piper, D. J.~W.,
  Preusser, F., Renssen, H., Roberts, D.~H., Roche, D.~M., Saint-Ange, F.,
  Stroeven, A.~P., Teller, J.~T., 2015. {On the reconstruction of palaeo-ice
  sheets: Recent advances and future challenges}. Quaternary Science Reviews
  125, 15--49.

\bibitem[{Unger(2014)}]{Unger:2014uv}
Unger, J.~D., 2014. {Regulating the Arctic Gold Rush: Recommended Regulatory
  Reforms to Protect Alaska's Arctic Environment from Offshore Oil Drilling
  Pollution }. Alaska L Rev 31.

\bibitem[{van~den Berk and Drijfhout(2014)}]{vandenBerk:2014cd}
van~den Berk, J., Drijfhout, S.~S., Sep. 2014. {A realistic freshwater forcing
  protocol for ocean-coupled climate models}. Ocean Modelling 81, 36--48.

\bibitem[{Vernet et~al.(2012)Vernet, Smith, Cefarelli, Helly, Kaufmann, Lin,
  Long, Murray, Robison, Ruhl, Shaw, Sherman, Sprintall, Stephenson, Stuart,
  and Twining}]{Vernet:2012th}
Vernet, M., Smith, K. L.~J., Cefarelli, A.~O., Helly, J.~J., Kaufmann, R.~S.,
  Lin, H., Long, D.~G., Murray, A.~E., Robison, B.~H., Ruhl, H.~A., Shaw,
  T.~J., Sherman, A.~D., Sprintall, J., Stephenson, G. R.~J., Stuart, K.~M.,
  Twining, B.~S., Sep. 2012. {Islands of Ice: Influence of Free-Drifting
  Antarctic Icebergs on Pelagic Marine Ecosystems}. Oceanography 25~(3),
  38--39.

\bibitem[{Wagner et~al.(2016)Wagner, Dell, and Eisenman}]{Wagner:2016up}
Wagner, T. J.~W., Dell, R.~W., Eisenman, I., Oct. 2016. {An analytical model of
  iceberg drift}. arXiv.org.

\bibitem[{Weeks and Mellor(1978)}]{Weeks:1978vi}
Weeks, W.~F., Mellor, M., Mar. 1978. {Some elements of iceberg technology}.
  Tech. rep., Cold Regions Research and Engineering Laboratory.

\bibitem[{Wiersma and Jongma(2010)}]{Wiersma:2010jj}
Wiersma, A.~P., Jongma, J.~I., Aug. 2010. {A role for icebergs in the 8.2 ka
  climate event}. Climate Dynamics 35~(2-3), 535--549.

\end{thebibliography}

\end{document}